\documentclass[%
reprint,
superscriptaddress,
amsmath,amssymb,amsthm,mathrsfs,
aps,
prl,
]{revtex4-2}

\usepackage{graphicx}% Include figure files
\usepackage{dcolumn}% Align table columns on decimal point
\usepackage{bm}% bold math
\usepackage[usenames]{color}
\usepackage[ %showframe,%Uncomment any one of the following lines to test
margin=0.75in,
]{geometry}
\usepackage[normalem]{ulem}

\graphicspath{{./figs/}}
\def\TN{$T_\mathrm{N}$}

\def\TC{$T_{\mathrm{C}}$}

\def\Rw{$R_w$}

\def\muSR{$\mu$SR}

\def\SrBaMnTi{Sr$_{1-x}$Ba$_x$Mn$_{1-y}$Ti$_{y}$O$_3$}
\def\SrBa{Sr$_{1-x}$Ba$_x$MnO$_3$}

\def\Srba{(Sr,Ba)MnO$_3$}
\def\SBMTO{(Sr,Ba)(Mn,Ti)O$_3$}
\def\sampleA{Sr$_{0.55}$Ba$_{0.45}$MnO$_3$}

\def\sampleB{Sr$_{0.4}$Ba$_{0.6}$Mn$_{0.95}$Ti$_{0.05}$O$_3$}

\begin{document}
	
\title{Local atomic and magnetic structure of multiferroic (Sr,Ba)(Mn,Ti)O$_3$}
	
	\author{Braedon Jones}
	\affiliation{ %
		Department of Physics and Astronomy, Brigham Young University, Provo, Utah 84602, USA.
	} %
	\author{Christiana Z. Suggs}
	\affiliation{ %
		Department of Physics and Astronomy, Brigham Young University, Provo, Utah 84602, USA.
	} %

         \author{Elena Krivyakina}
        \affiliation{
        Department of Physics, Northern Illinois University, DeKalb, Illinois 60115, USA.
        }
        \affiliation{
         Materials Science Division, Argonne National Laboratory, Lemont, Illinois 60439, USA.
        }

        \author{Daniel Phelan}
        \affiliation{
        Materials Science Division, Argonne National Laboratory, Lemont, Illinois 60439, USA.
        }

        \author{V. Ovidiu Garlea} 
        \affiliation{%
        Neutron Scattering Division, Oak Ridge National Laboratory, Oak Ridge, Tennessee 37831, USA.
        }%
        
        \author{Omar Chmaissem}
        \affiliation{
        Department of Physics, Northern Illinois University, DeKalb, Illinois 60115, USA.
        }
        \affiliation{
         Materials Science Division, Argonne National Laboratory, Lemont, Illinois 60439, USA.
        }

	\author{Benjamin A. Frandsen}
	\affiliation{ %
		Department of Physics and Astronomy, Brigham Young University, Provo, Utah 84602, USA.
	} %

\begin{abstract}
    We present a detailed study of the local atomic and magnetic structure of the type-I multiferroic perovskite system (Sr,Ba)(Mn,Ti)O$_3$ using x-ray and neutron pair distribution function (PDF) analysis, polarized neutron scattering, and muon spin relaxation ($\mu$SR) techniques. The atomic PDF analysis reveals widespread nanoscale tetragonal distortions of the crystal structure even in the paraelectric phase with average cubic symmetry, corresponding to incipient ferroelectricity in the local structure. Magnetic PDF analysis, polarized neutron scattering, and \muSR\ likewise confirm the presence of short-range antiferromagnetic correlations in the paramagnetic state, which grow in magnitude as the temperature approaches the magnetic transition. We show that these short-range magnetic correlations coincide with a reduction of the tetragonal (i.e. ferroelectric) distortion in the average structure, suggesting that short-range magnetism can play an important role in magnetoelectric and/or magnetostructural phenomena even without genuine long-range magnetic order. The reduction of the tetragonal distortion scales linearly with the local magnetic order parameter, pointing to spontaneous linear magnetoelectric coupling in this system. These findings provide greater insight into the multiferroic properties of (Sr,Ba)(Mn,Ti)O$_3$ and demonstrate the importance of investigating the local atomic and magnetic structure to gain a deeper understanding of the intertwined degrees of freedom in multiferroics.
    
    %The average structure of \SrBa\ has been previously examined to better understand its antiferromagnetic and antiferroelectric phase transitions. However, local analysis through Neutron, X-ray, and Muon diffraction was conducted to reveal the local structure evolution as the material experiences phase transitions imposed through varying temperatures. This study shows, through PDF analysis, that there are short-range tetragonal distortions above the antiferromagnetic transition temperature. Moreover, magnetic correlations develop 100K above the Neel temperature influencing the long-range structure of \SrBa.
\end{abstract}
	
\maketitle

\section{Introduction}
% Multiferroic materials exhibit more than one primary ferroic order simultaneously~\cite{schmi;f94}. The focus here is on magnetoelectric multiferroics, which have simultaneously ordered magnetic moments (ferromagnetic or antiferromagnetic) and electric dipoles (ferroelectric or antiferroelectric). The coexistence of (anti)ferromagnetic and (anti)ferroelectric orders enables coupling between the two, and therefore the possibility for electric control of magnetism and vice versa. Such a scenario has myriad potential applications in information science, energy transformation, signal processing, and more. It also creates a platform for fundamental study of entangled orders in complex materials~\cite{rames;nm07,dong;advp15,fiebi;nrm16,spald;nm19}. Local symmetry breaking and short-range magnetic correlations are vital for a full microscopic understanding of the complex phase behavior in multiferroics~\cite{skjae;prx19}, yet very little work has been done along these lines. The initial focus will be on (Sr,Ba)MnO$_3$, a type-I multiferroic with an unusually high magnetic ordering temperature, and TbMnO$_3$, a classic type-II multiferroic. PDF experiments will reveal precisely how the local and average structure (which set the polarization) change in response to the development of short- and long-range magnetic correlation. In-situ magnetic fields will be used to probe the magnetoelectric coupling directly. These results will guide follow-up work on doped variants and related materials.

In magnetoelectric multiferroic materials, ordered magnetic moments exist simultaneously with ordered electric dipoles~\cite{schmi;f94}. The coexistence of magnetic and electric orders enables coupling between the two, and therefore the possibility for electric control of magnetism and vice versa. Such a scenario has myriad potential applications in information science, energy transformation, signal processing, and more~\cite{vopso;crssms15}, while also creating a platform for fundamental study of entangled orders in complex materials~\cite{rames;nm07,dong;advp15,fiebi;nrm16,spald;nm19}.

%Multiferroics can be classified as type I or type II~\cite{schmi;f94}. In type-I multiferroics, the ferroelectric and magnetic transitions are typically widely separated in temperature and fairly independent of each other, as in BiFeO$_3$. These materials frequently offer technologically favorable polarization magnitudes and ordering temperatures, but cross-order control is limited due to the weak coupling between the orders. Type-II multiferroics such as TbMnO$_3$ are defined by a magnetically-driven ferroelectric transition, so the two orders are very tightly linked, but the polarization is often much weaker and occurs at much lower temperatures, limiting the practical impact. A major goal in the field is to discover or engineer materials that possess all of the desirable properties for technological application~\cite{spald;nm19}. A prerequisite for this goal is obtaining a more complete understanding of the details of the magnetoelectric coupling in representative systems.

%In spite of the potential use of multiferroics, they are a rare class of materials, as it is common for the mechanisms that cause the magnetic and electric order to reduce the existence of each other~\cite{kimur;prb03}. For instance, electric dipole moments can be caused by,,,, which actually diminishes the magnetism as ,,,,.

Multiferroics are typically classified as type I or type II~\cite{schmi;f94}. In type-I multiferroics, the ferroelectric and magnetic transitions are usually widely separated in temperature and fairly independent of each other. These materials frequently offer technologically favorable electric polarization magnitudes and ordering temperatures, but cross-order control is limited due to the weak coupling between the orders. Type-II multiferroics are defined by a magnetically-driven ferroelectric transition, such that the two orders are very tightly linked. However, the polarization is often much weaker and restricted to low temperature, thereby limiting the potential for practical application. A major goal in the field is to discover or engineer materials that possess all of the desirable properties for technological application~\cite{spald;nm19}. A prerequisite for this goal is obtaining a more complete understanding of the details of the magnetoelectric coupling in representative systems.

In this context, the perovskite material \Srba\ has recently attracted interest as a promising multiferroic system~\cite{chmai;prb01, sakai;prl11, pratt;prb14, goian;jpcm16, somai;prm18, chapa;prm19}. It is a type-I multiferroic with a relatively large electric polarization of $\sim$4.5~$\mu$C/cm$^2$, yet the ferroelectric (FE) and antiferromagnetic (AFM) orders are strongly coupled in a manner similar to prototypical type-II multiferroics. This unusually strong magnetoelectric coupling is understood as a consequence of the Mn sublattice being responsible for both the electric polarization and the magnetic order, providing a unique route to outstanding multiferroic properties in \Srba~\cite{sakai;prl11}.

%\begin{figure}
%	\includegraphics[width=65mm]{figs/Better figs/Molecular Structure Labeled.png}
%	\caption{\label{fig:struc} Cubic lattice of \Srba from Vesta. The center green atomic site represents either Mn or Ti and is the site of the cause of both the antiferromagnetic and ferroelectric ordering. The purple spheres represent Sr or Ba, while the red spheres represent the Oxygen sites. With different temperatures, the structure evolves from Cubic to Tetragonal}
%\end{figure}

In the paraelectric state, \Srba\ adopts the typical cubic perovskite structure. The FE transition, which occurs between 330 and 430~K depending on the composition, is associated with an off-centering displacement of the Mn$^{4+}$ ion relative to the surrounding O$^{2-}$ ions and a tetragonal distortion, in which $c$ becomes slightly greater than $a$. The resulting space group is $P4mm$~\cite{somai;prm18}. Below the AFM N\'eel temperature \TN, which ranges from 130 to 200~K depending on the composition, G-type AFM order develops. The magnetic structure is characterized by anti-parallel alignment between nearest-neighbor Mn$^{4+}$ ions. The onset of magnetic order coincides with a significant reduction of the tetragonal (and therefore ferroelectric) distortion to about 30\% of the maximum distortion, demonstrating strong magnetoelectric coupling. Interestingly, this reduction of the FE distortion begins nearly 100~K above \TN\ and grows as the temperature approaches the magnetic transition. It has been speculated, though not shown, that short-range magnetic correlations in the paramagnetic state are responsible for this behavior~\cite{sakai;prl11, pratt;prb14}. 

\begin{figure}
    \includegraphics[width=80mm]{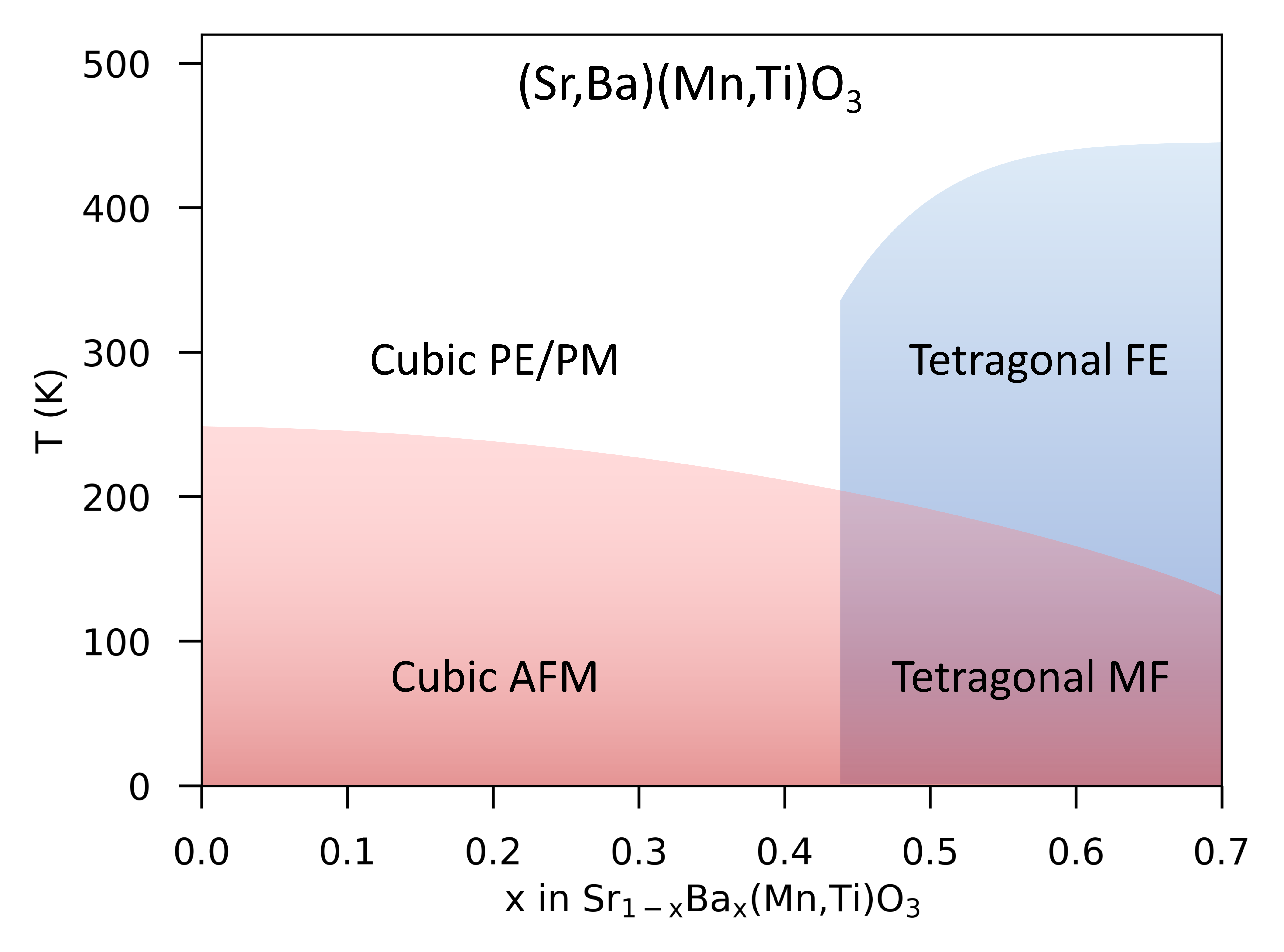}
    \caption{Phase diagram for \SrBaMnTi\ as a function of temperature and barium content $x$. Compositions with $x \ge 0.5$ are synthesized with dilute amounts of Ti on the Mn site. Compiled from results published in Refs.~\onlinecite{sakai;prl11,somai;prm18,chapa;prm19}. PE = paraelectric; PM = paramagnetic; FE = ferroelectric; AFM = antiferromagnetic; MF = multiferroic.}
    \label{fig:phasediagram}
\end{figure}

The FE and AFM ordering temperatures can be tuned systematically by varying the Ba content in \SrBa. Ferroelectricity first appears for $x=0.43$ with a Curie temperature of $T_{\mathrm{C}}=330$~K, which rises rapidly to about 400~K for $x=0.5$. Compositions with $x\ge0.5$ can be stabilized by co-substitution with small amounts of Ti on the Mn site~\cite{chapa;prm19}, with a maximum Ba content of $x=0.7$ achievable with a Ti content of $y=0.12$ in \SrBaMnTi. \TC\ increases slightly to approximately 430~K for the largest Ba concentrations. Meanwhile, the N\'eel temperature \TN\ decreases from 200~K for $x=0.43$ to approximately 130~K for $x=0.7$~\cite{chapa;prm19}. The phase diagram in the space of temperature versus Ba content is given in Fig.~\ref{fig:phasediagram}.

Features of the local structure such as short-range structural correlations that differ from the long-range crystallographic structure have long been recognized as crucial for understanding ferroelectric materials. For example, short-range tetragonal distortions resulting in local polar displacements within an average centrosymmetric crystal structure have been observed in numerous perovskite-based ferroelectrics ranging from classic systems such as BaTiO$_3$ to complex relaxor ferroelectrics~\cite{kwei;ferro95,egami;armr07,hou;jecs18,li;afm18,yoned;jsapr23}. The presence and evolution of such short-range ferroelectric distortions have yielded insights into the transition mechanism and origin of the observed ferroelectric behavior. Many of these local-structure studies have utilized pair distribution function (PDF) analysis of x-ray and neutron total scattering data, one of the most powerful methods available for quantitatively elucidating local atomic structure~\cite{egami;b;utbp12}. By Fourier transforming the total scattering signal, i.e. both the Bragg peaks arising from the average crystal structure and the diffuse scattering from local deviations from the average structure, the PDF provides a real-space view of the local structure in the form of a histogram of interatomic distances on length scales ranging from the nearest-neighbor distances to several nanometers or longer. The method has had a major impact on structural studies of functional materials, quantum materials, and more~\cite{young;jmc11,zhu;advs21,akamo;ejm22}.

Studies of local atomic structure are likewise expected to be valuable for investigations of multiferroics, especially when combined with probes of the local \textit{magnetic} structure. Detailed understanding of short-range atomic and magnetic correlations could potentially provide a direct view into the microscopic mechanisms of magnetoelectric coupling in multiferroics. Several fruitful atomic PDF studies have been conducted on multiferroic systems~\cite{kodam;jpsj07, jeong;prb11, tripa;prb12, sim;prb14, jiang;jssc17, skjae;prx19}, but without corresponding information regarding the local magnetic correlations, resulting  in an incomplete picture of the local electronic environment. Magnetic PDF (mPDF) analysis can complete the picture by providing details of local magnetism. Similar to atomic PDF, the mPDF is obtained by Fourier transforming the magnetic total scattering signal, yielding the pairwise magnetic correlations in real space~\cite{frand;aca14,frand;aca15}. Standard neutron total scattering measurements of magnetic materials produce both the atomic and magnetic PDF together in a single total PDF pattern, allowing for simultaneous investigation of the local atomic and magnetic structures together. Joint atomic and magnetic PDF studies have been successfully conducted for a variety of different magnetic materials~\cite{frand;prl16,frand;prb16,frand;prm17,frand;prb20,fletc;prb21,ander;ij21,baral;matter22,baral;afm23}, but this approach has not been applied to multiferroics up to this point.

Here, we report a thorough x-ray and neutron total scattering study of \SrBaMnTi\ for $0.45 \le x \le 0.7$  and $0.00 \le y \le 0.12$ featuring combined atomic and magnetic PDF analyses of the local atomic and magnetic structure. We also performed complementary muon spin relaxation (\muSR) and polarized neutron diffraction experiments to further probe the local magnetism. We show that short-range ferroelectric distortions with a typical length scale of $\sim$2~nm are present in the instantaneous local structure above \TC\ in all measured compounds. Short-range antiferromagnetic correlations likewise appear in the ferroelectric state about 100~K above the magnetic ordering temperature and grow in magnitude as the temperature approaches \TN. The growth of the local magnetic order parameter scales linearly with a reduction of the tetragonal distortion in the crystal structure above \TN, indicating that the magnetoelectric coupling mechanism can have effect even in the absence of genuine long-range magnetic order when robust short-range magnetic correlations are present.

\section{Methods}
High-purity powder samples of \SrBaMnTi\ were synthesized according to previously published recipes~\cite{somai;prm18,chapa;prm19}. The compositions prepared for the current study are $(x, y) = (0.43,0), (0.45, 0), (0.6, 0.05)$, and $(0.7, 0.12)$. The phase purity and composition of each sample were verified by x-ray diffraction.

\begin{figure}
	\includegraphics[width=80mm]{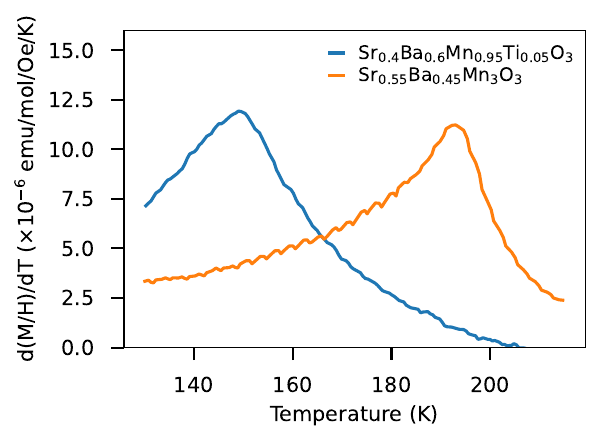}
	\caption{\label{fig:magnetometry} Temperature derivative of the magnetic susceptibility of \sampleB\ and \sampleA\ as a function of temperature. Maxima occur at 149~K and 193~K, respectively. The measurements were taken in a field of 1~T using a zero-field-cooling protocol.}
    \centering
\end{figure}

Magnetic susceptibility measurements ($M/H$) were performed using a Quantum Design Magnetic Property Measurement System (MPMS-3). The data were collected in a field of 1~T upon warming after the samples were cooled to base temperature in zero magnetic field (ZFC-W). We show in Fig.~\ref{fig:magnetometry} the temperature derivative of the magnetic susceptibility for \sampleA\ and \sampleB. \TN\ as determined by the maximum of the curve is 193~K for \sampleA\ and 149~K for \sampleB.
These values are consistent with the results published in Ref.~\onlinecite{chapa;prm19}, confirming the expected magnetic behavior for these samples. We note that the ferroelectric behavior of samples of similar composition prepared by identical methods was confirmed in Ref.~\onlinecite{goian;jpcm16}.

Neutron total scattering experiments were performed on the NOMAD beamline~\cite{neuef;nimb12} at the Spallation Neutron Source (SNS) at Oak Ridge National Laboratory. Samples of mass $\sim$200~mg were loaded into thin quartz capillaries and mounted on the beamline. Diffraction patterns with a total integrated proton charge of 4~C were collected at a series of temperatures between 90 and 500~K using a nitrogen cryostream to control the temperature. The time-of-flight scattering data were reduced and Fourier transformed with $Q_{\mathrm{max}}=30$~\AA$^{-1}$ to produce the PDF data using the ADDIE software~\cite{mcdon;aca17}. X-ray total scattering measurements were performed on beamline 28-ID-1 at the National Synchotron Light Source II (NSLS-II) at Brookhaven National Laboratory. Samples of mass $\sim$15~mg were loaded into polyimide capillaries placed in a liquid helium cryostat, and mounted on the beamline. We collected scattering patterns between 250~K and 500~K in steps of 5~K. The resulting diffraction patterns were azimuthally integrated using pyFAI~\cite{kieff;jpconfs13} and normalized and Fourier transformed into the PDF data using xPDFsuite~\cite{yang;arxiv15}. Models of the atomic structure were fit to the neutron and x-ray PDF data using PDFgui~\cite{farro;jpcm07}. Magnetic PDF fits were carried out using the \texttt{diffpy.mpdf} package~\cite{frand;jac22}. We note that the total scattering data recorded in both the neutron and x-ray experiments contain scattering intensity integrated over all energy transfers, and the PDF therefore corresponds to the instantaneous local structure of the material.

Polarized neutron scattering measurements were performed on the HYSPEC instrument at SNS~\cite{zaliz;jpconfs17}. The spin-flip (SF) and non-spin-flip (NSF) cross sections were measured to enable the separation of the magnetic scattering from the nuclear and nuclear spin-incoherent scattering cross sections, as explained in detail elsewhere~\cite{schar;pssa93,stewa;jac09,ehler;rsi13}. The energy of the incident neutrons was $E_i=28$~meV, and the frequency of the Fermi chopper was 60~Hz. The data were integrated over energy transfers from $-3$ to 3~meV. SF and NSF data were collected with the neutron polarization oriented parallel to the $Q$ vector corresponding to the elastic scattering at the center  of the 60$^{\circ}$ detector bank, for six different detector positions. The total  $2\theta$ scattering range covered by the measurements was 4 - 111$^{\circ}$. The flipping ratio was found to be 14 through measurements of the nuclear Bragg peaks in both the SF and NSF channels. MANTID was used to implement post-processing data corrections for the angle-dependent supermirror transmission and flipping ratio efficiency~\cite{zaliz;jpconfs17,savic;jpconfs17}. In this study, the magnetic scattering is determined as the intensity measured in the SF channel corrected for the flipping ratio using Eq.~2 in Ref.~\onlinecite{zaliz;jpconfs17}.

Muon spin relaxation (\muSR) experiments were conducted at the TRIUMF Laboratory in Vancouver, Canada using the LAMPF spectrometer on the M20D beamline. In a \muSR\ experiment, spin-polarized positive muons are implanted in the sample, where they undergo Larmor precession in the local magnetic field (i.e. the vector sum of any internal and externally applied fields) at their stopping sites. After a mean lifetime of 2.2~$\mu$s, the muons undergo asymmetric decay into positrons, which are emitted preferentially along the direction of the muon spin at the moment of decay~\cite{hilli;nrmp22}. The time-dependent asymmetry, $a(t)$, is measured as the difference in positron events between two detectors placed on opposite sides of the sample as a function of time after muon implantation. This quantity is proportional to the projection of the net muon spin polarization along the direction connecting the two detectors. The behavior of $a(t)$ as a function of temperature provides information about the evolution of the local magnetic field distribution across the magnetic phase transition. Importantly, \muSR\ is sensitive both to long- and short-range magnetic correlations, which can be either static or dynamically fluctuating. We performed the \muSR\ data analysis using the open source program BEAMS~\cite{peter;gh21}. 

\section{Results}

\subsection{Characterization of the local atomic structure}
We first present results relating to the local atomic structure of \SBMTO\ determined from atomic PDF analysis of the neutron and x-ray total scattering data. The focus is on the specific compositions \sampleA\ and \sampleB. We begin with model-independent analysis of the data, which we then support through quantitative refinements of structural models.

\subsubsection{Model-independent analysis} 
\begin{figure*}
	\includegraphics[width=170mm]{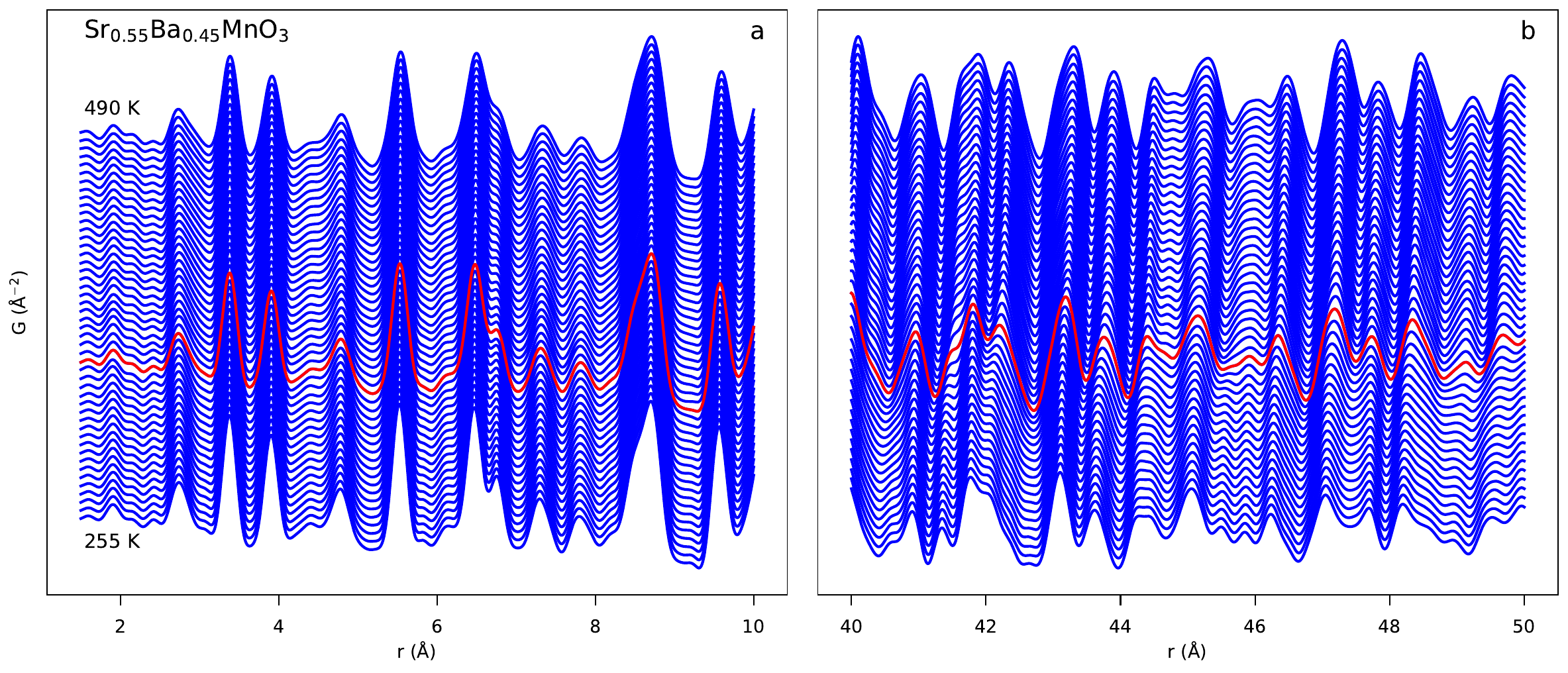}
	\caption{\label{fig:WaterFallPlot} X-ray PDF data for \sampleA\ between 250 K and 490 K displayed over the range 1.5 -- 10~\AA\ (a) and 40 -- 50~\AA\ (b). The PDF patterns were collected in increments of 5~K and are offset vertically on the plot for clarity. For the viewing range at higher $r$, clear and abrupt changes are evident across the ferroelectric transition at 350~K (red PDF pattern). No abrupt changes are observed in the low-$r$ viewing range.}
    \centering
\end{figure*}

To illustrate the temperature dependence of the structure on a qualitative basis, we display in Fig.~\ref{fig:WaterFallPlot} the PDF data for \sampleA\ at various temperatures across the ferroelectric transition (\TC~= ~350~K). Two data ranges are shown: 1.5 -- 10~\AA, revealing the local structure, and 40 --- 50~\AA, corresponding to intermediate-range structural correlations. The PDF patterns remain nearly unchanged with temperature over the shorter range, with only a slight broadening and shifting of the peaks to higher $r$ visible as the temperature is raised from 255~K to 490~K. No abrupt change occurs across \TC\ (highlighted by the red PDF curve). In contrast, clear and abrupt changes are seen across \TC\ when inspecting the PDF patterns over the longer data range, such as a rightward shift of the shoulder peak centered around 42~\AA, a merging of the small peaks between 45.5~\AA\ and 46~\AA, and many other systematic changes as the temperature increases across \TC. Thus, the structural change corresponding to the ferroelectric transition at \TC\ is clearly evident in the data over sufficiently long length scales such as 40 -- 50~\AA, while the local structure shows no such change. We can therefore infer that the instantaneous local structure is already symmetry-broken well above \TC\ due to short-range ferroelectric distortions.

\begin{figure*}
	\includegraphics[width=170mm]{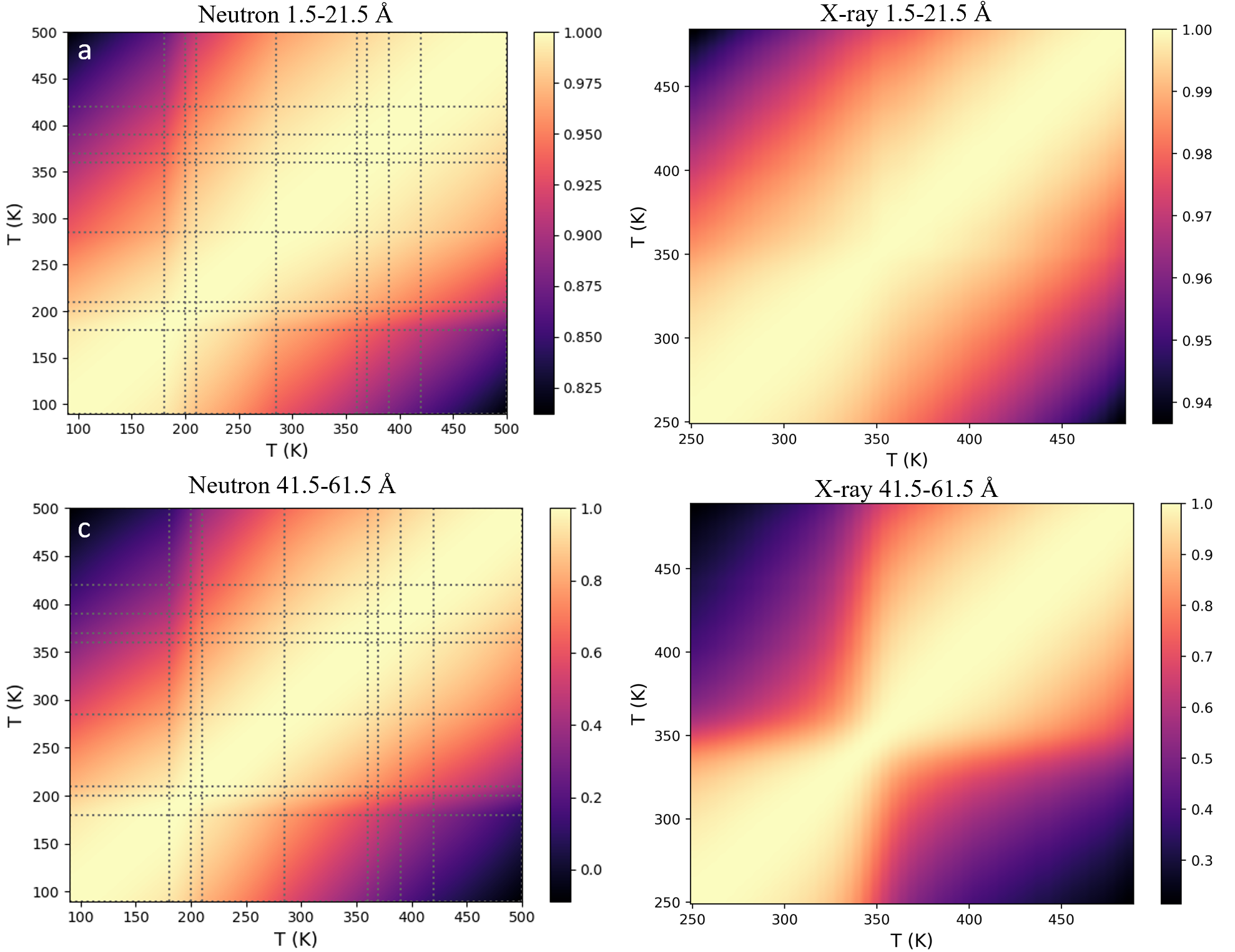}
	\caption{\label{fig:pearsonmap} Color maps of the Pearson correlation coefficient for pairs of PDF data sets obtained from \sampleA\ at various temperatures. (a, b) Correlation coefficients computed for PDF data in the range 1.5 -- 21.5~\AA\ using neutrons and x-rays, respectively. Note the difference in temperaure ranges for the two types of experiments. (c, d) Same as (a, b), except that the data range 41.5 -- 61.5~\AA\ was used. Vertical and horizontal dashed lines in (a) and (c) mark the temperatures at which neutron PDF data were collected. X-ray data were collected on a uniform temperature grid between 250~K and 500~K in steps of 5~K.}
 %combine the two color maps, center maybe, make r mid bigger and produce a better title 
\end{figure*}

To investigate this further, we computed the Pearson correlation coefficient between each PDF pattern and the patterns collected at every other temperature for a given sample. This quantity, given by 
\begin{equation}
    \label{eq:pearson}
     R = \frac{\sum_i(G_{1,i}-\langle G_1\rangle)(G_{2,i}-\langle G_2\rangle)}{\sqrt{\sum_i(G_{1,i}-\langle G_1\rangle)^2\sum_i(G_{2,i}-\langle G_2\rangle)^2}},
\end{equation}
can be used as a measure of the similarity between two data sets $G_1$ and $G_2$, with angular brackets denoting the mean and the subscript $i$ indexing each discrete data point. The Pearson coefficient ranges from 1 to $-1$, with 1 indicating perfect linear correlation, 0 indicating no correlation, and $-1$ indicating anti-correlation between the two data sets. We computed the coefficient for the data ranges 1.5 -- 21.5~\AA\ and 41.5 -- 61.5~\AA\ to allow a comparison of the local and longer-range structure once again. The resulting Pearson correlation maps are shown for \sampleA\ in Fig.~\ref{fig:pearsonmap} using both neutron and x-ray PDF data. These plots can be read by selecting a pair of temperatures whose data sets are to be compared, locating the intersection of these two temperature points on the plot, and then comparing the observed color at that point to the color bar shown to the right of each plot. The main diagonal from lower left to upper right always has a coefficient of unity (since the diagonal corresponds to each data set being compared to itself), and the maps are symmetrical when reflected across the main diagonal. Bright colors indicate a high level of similarity between PDF data collected at the corresponding temperatures, while dark colors indicate less similarity. % Panels (a) and (c) in Fig.~\ref{fig:pearsonmap} show the correlations calculated from the neutron PDF data, which were collected at 10 temperatures between 90 and 500~K indicated by the dashed gray lines. Panels (b) and (c) show the corresponding x-ray PDF results using data collected in steps of 5~K between 5 and 500~K. The top and bottom panels show the correlation coefficients calculated from the shorter and longer data ranges, respectively. 

Several features of the correlation plots bear mentioning. As can be seen by the color scales, the overall variation across the full temperature range is significantly less for the shorter data range than for the longer data range, consistent with our inspection of the PDF data in Fig.~\ref{fig:WaterFallPlot}. The ferroelectric and antiferromagnetic transitions are identifiable in the longer data range as relatively sharp changes in the correlation coefficient in certain regions of the correlation maps. Specifically, a rapid variation is visible in the long-range x-ray data [Fig.~\ref{fig:pearsonmap}(d)] when traversing 350~K, corresponding to the expected \TC\ of this composition. This indicates that the structure on the length scale of 41.5 -- 61.5~\AA\ changes significantly and abruptly across 350~K. In contrast, the variation across 350~K for the short-range data [Fig.~\ref{fig:pearsonmap}(b)] is much less pronounced, again supporting the idea that the local structure is already symmetry-broken above \TC\ and does not undergo any significant rearrangement at the long-range ferroelectric transition. Evidence of a structural change is less visible for the long-range neutron data [Fig.~\ref{fig:pearsonmap}(c)], which we attribute primarily to sparser temperature coverage across \TC\ compared to the x-ray data. On the other hand, the antiferromagnetic transition can be seen in both the short- and long-range neutron data [Fig.~\ref{fig:pearsonmap}(a, c)] around 200~K, the expected \TN\ for this composition. %The x-ray results also show a variation in the correlation coefficients across $\sim$120~K, albeit much less well-defined than the transition visible at 350~K. We attribute this feature at 120~K to the antiferromagnetic transition. The discrepancy with the expected N\'eel temperature of 200~K is due to known issues with low-temperature equilibration for insulating samples in the cryostat used on the x-ray PDF beamline; in other words, although the nominal cryostat temperature may be 120~K, the actual sample temperature could be much higher, such as 200~K. %The final comment we make on these correlation maps is the fact that the antiferromagnetic transition is roughly as evident in the shorter data range as in the longer data range. This indicates that the local structure behaves similarly to the longer-range structure across the antiferromagnetic transition, in contrast to the significant difference between the local and long-range structure across the ferroelectric transition.

Some caution must be exercised when basing a claim of local symmetry breaking (such as the one we made in the previous discussion) on the observation that the low-$r$ region of the PDF shows a higher Pearson correlation as a function of temperature than high-$r$ regions. The magnitude of the shifts in PDF peak positions due to thermal expansion scales proportionally with $r$. Therefore, even for a conventional material in which the local structure shows no deviations from the average structure, correlation maps such as the ones displayed in Fig.~\ref{fig:pearsonmap} will always show higher correlations for shorter data ranges than for longer data ranges. To verify that the correlation maps generated from our PDF data actually contain evidence for local symmetry breaking rather than merely thermal expansion, we corrected the computed Pearson correlations for the effects of thermal expansion as follows.

First, we performed fits in PDFgui to representative low-temperature (250~K), intermediate-temperature (375~K), and high-temperature (500~K) data sets. We then simulated the PDF pattern at 250~K and 500~K using the lattice parameters refined at those temperatures but all other structural parameters (atomic coordinates, atomic displacement parameters, and the linear sharpening parameter in PDFgui) as determined by the refinement at the intermediate temperature. Thus, the simulated 250~K pattern differs from the simulated 500~K pattern solely due to the thermal shifts of the lattice parameters. The Pearson correlation coefficient between these two simulated PDF patterns was then calculated over the desired data range. We will call this coefficient $R_{\mathrm{th}}$. The quantity $\Delta R_{\mathrm{th}} = 1 - R_{\mathrm{th}}$ then represents the contribution of thermal expansion to the differences between the data sets collected at 90 and 500~K. By adding $\Delta R_{\mathrm{th}}$ to the experimental Pearson correlation coefficient $R_{\mathrm{exp}}$ computed from the actual data sets, we obtain a corrected correlation coefficient $R_{\mathrm{c}}=R_{\mathrm{exp}}+\Delta R_{\mathrm{th}}$ for which the influence of thermal expansion has been removed (albeit crudely). To apply this correction to the correlation coefficients calculated for all other temperatures with respect to 90~K, the correction term $\Delta R_{\mathrm{th}}$ was scaled linearly from 0 at 90~K to its maximum value at 500~K, which presupposes linear thermal expansion.

\begin{figure}
	\includegraphics[width=75mm]{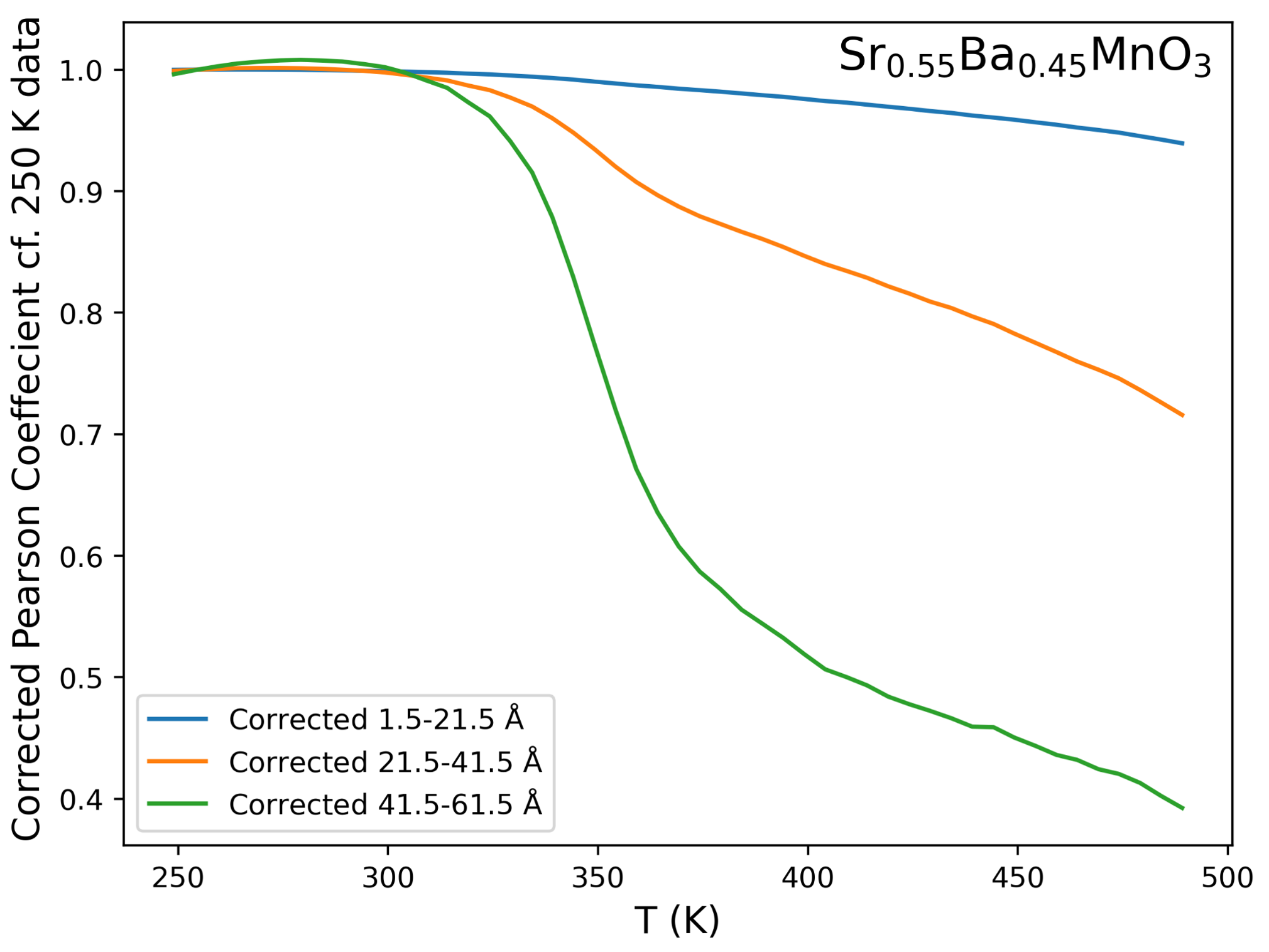}
	\caption{\label{fig:PearsonCutout} Pearson correlation coefficient for \sampleA\ with respect to the x-ray PDF data collected at 250 K, corrected for thermal expansion as explained in the main text. Three different data ranges are used. The ferroelectric transition around 350~K is most evident for the highest $r$-range but essentially invisible for the shortest data range, confirming that the local structure is already symmetry broken above \TC.}
 
\end{figure}

The temperature dependence of the corrected Pearson correlation coefficient with respect to the 250~K x-ray PDF data for \sampleA\ is displayed for three different data ranges in Fig.~\ref{fig:PearsonCutout}.
As the temperature increases toward and across \TC, the shorter data ranges show higher correlation coefficients (both corrected and uncorrected) than the long data range. This further supports the idea that the local structure undergoes no sharp change across \TC\ because the local symmetry is already broken in the paraelectric state. The transition appears as step-like change around 350~K in the correlation curve for the 41.5 -- 61.5~\AA\ data range, with a slightly broader step observed for 21.5 --41.5~\AA. Similar results are found for the neutron PDF data and for the other compositions.

\subsubsection{PDF boxcar fits}

Complementing the model-independent analysis presented in the previous section, we now present structural refinements of the tetragonal model against the PDF data. We used a boxcar fitting approach, in which the structural model was refined against a sliding window through the PDF data. In our case, we performed fits over a 20~\AA\ range beginning with 1.5 -- 21.5~\AA\ and ending with 41.5 --61.5~\AA\ in increments of 2~\AA, resulting in 21 distinct fitting ranges for each temperature and hundreds of fits for each sample studied. This enables a detailed study of the evolution of the local structure (as determined from the shorter range fits) to the intermediate or average structure (as determined from the longer range fits). We used the tetragonal model at all temperatures, including those in the nominally cubic phase, to avoid imposing the symmetry of the average structure on the local structure. In the case that the structure really is cubic, the tetragonal lattice parameters will simply converge to equal values in the fit.  
% (this is good stuff, but it doesn't belong at this place in the manuscript) Pair Distribution Function analysis and fitting was performed via PDFgui and diffpy programs. PDF yields a weighted probability of a pair of atoms begin separated at a certain distance~\cite{young;jmc11}. In this process, reduced and Fourier Transformed data is fitted with least-squares regression through refinement of the lattice parameters, such as the lattice lengths, lattice positions, and anisotropic atom displacements (ADP).

\begin{figure}
	\includegraphics[width=75mm]{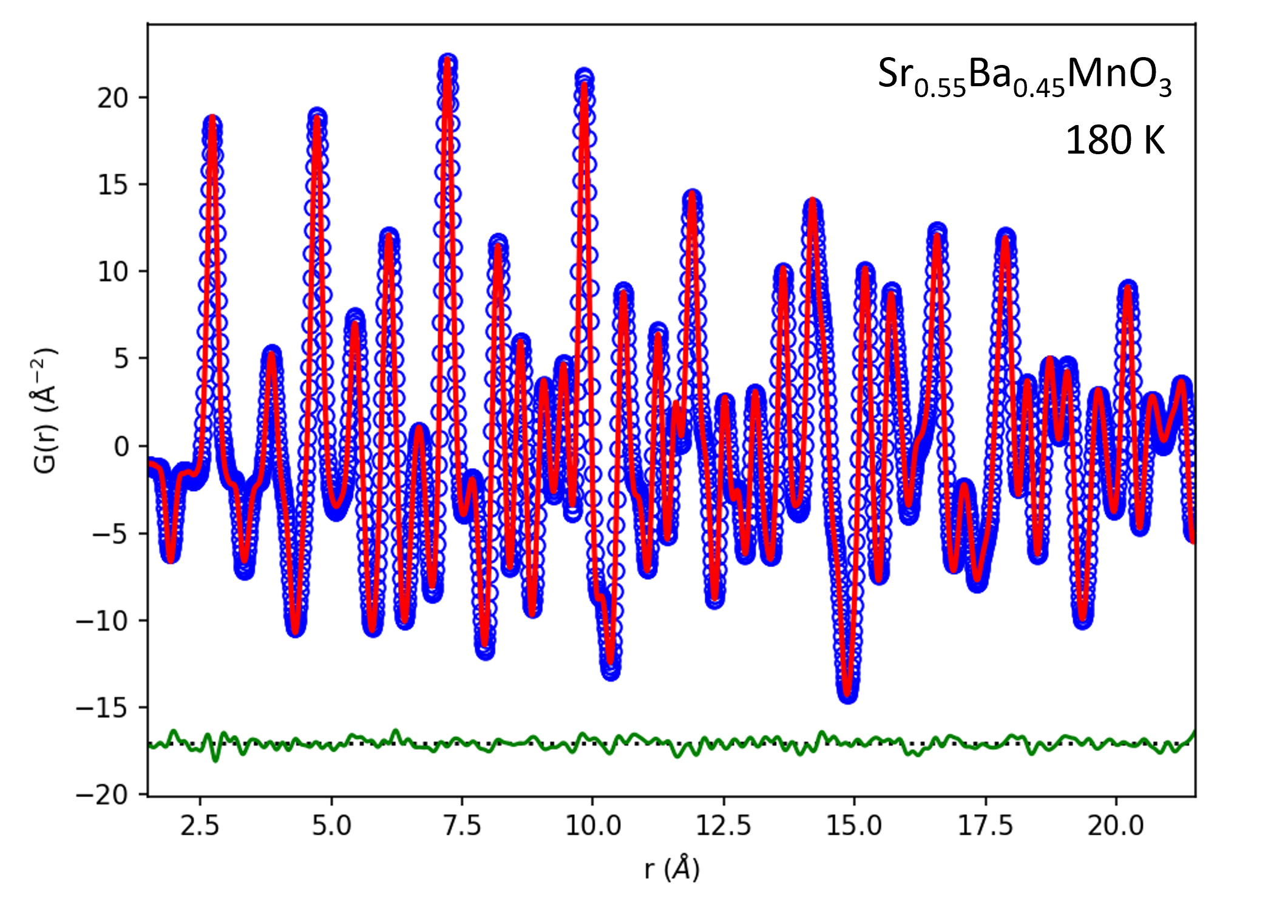}
	\caption{\label{fig:PDFfit} Representative neutron PDF fit for \sampleA\ at 180~K. The blue symbols show the experimental data, the red curve the best fit, and the green curve the fit residual, offset vertically for clarity.}
\end{figure}

The structural parameters refined in PDFgui included a scale factor, the lattice parameters, the coordinates of the atoms with symmetry-allowed positional degrees of freedom, the ADPs, and the linear correlated motion sharpening parameter for the fits that included data below 5~\AA. The $z$ coordinates of the atoms on the Mn site were fixed so that the positions of the other atoms could shift relative to the Mn sites. The goodness-of-fit metric \Rw\ typically lay between 0.05 to 0.10, indicating good quality fits. A representative fit is shown in Fig.~\ref{fig:PDFfit}, obtained from neutron PDF data for \sampleA\ at 180~K.

%(again, good stuff but not for here) This can be done locally because PDF utilizes both the Bragg diffraction and the diffuse scattering. The diffuse scattering informs of two-body correlations and thus more of the deviations found of the local structure~\cite{young;jmc11, proffen;zfk01}. An example PDF fitting is shown in in Fig.~\ref{fig:PDF analysis Fit}. The r range, in Angstroms, ranges from 1.5 to 21.5. These values are indicative of a local structure as opposed to average structure. Plotted on the y-axis is G(r) with units of inverse Angstroms squared. The intensity shows the probable expectation of of finding an atom at a certain distance away from a base atom. The blue data points are the resulting processed and Fourier transformed data from Beamline experiments and the red line is the fitting as produced from diffpy and PDFgui programming. The green line at the bottom graph is the difference between the observed and calculated data.

\begin{figure}
	\includegraphics[width=75mm]{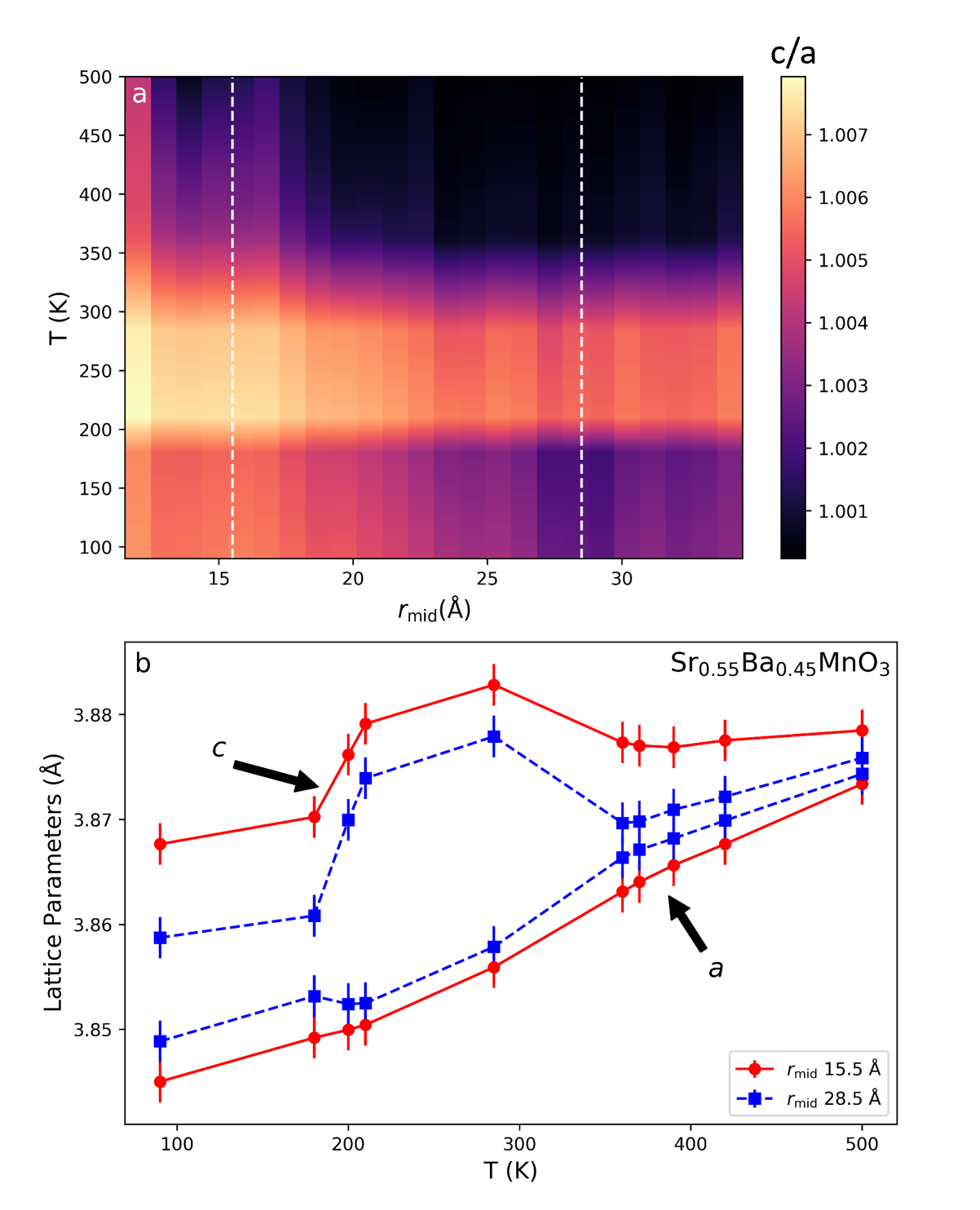}
	\caption{\label{fig:Ba45colormap} (a) Tetragonal distortion $c/a$ in \sampleA\ as a function of temperature and fitting range, as determined by PDF boxcar fits to the neutron PDF data. The horizontal axis indicates the midpoint of the 20-\AA\ fitting range $r_{\mathrm{mid}}$. Brighter (darker) colors correspond to a greater (lesser) tetragonal distortion. Linear interpolation was used between discrete $(r_{\mathrm{mid}}, T)$ points. (b) Lattice parameters extracted from the fits with $r_{\mathrm{mid}} = 15.5$~\AA\ and 28.5~\AA, corresponding to the vertical dashed lines in (a).}
\end{figure}

The refined lattice parameters were extracted from each fit conducted for a given sample. This allows us to quantify the temperature- and length-dependent behavior of the tetragonal (i.e. ferroelectric) distortion by calculating $c/a$ for every fit. The results for \sampleA\ based on neutron PDF fits are shown in Fig.~\ref{fig:Ba45colormap}(a).  
The globally ferroelectric phase is seen as the bright, horizontal band beginning at 350~K and extending to 200~K, below which AFM order sets in. We see that the the tetragonal distortion persists for all fitting ranges (albeit with a diminished magnitude) in the AFM phase, thereby confirming the multiferroic nature of the system. The reduced tetragonal distortion in the multiferroic phase is consistent with previous reports~\cite{sakai;prl11, somai;prm18}. At temperatures above 350~K, the long-range fits converge to $c/a=1$, evident as the dark region in the upper right corner of the plot. This is consistent with global cubic symmetry in the paralectric state. In contrast, the short-range fits with $r_{\mathrm{mid}}\lesssim 20$~\AA\ yield $c/a$ values that are slightly larger than unity and decrease with increasing fitting range, directly revealing a local tetragonal distortion with a length scale of approximately 2~nm. Vertical cuts along $r_{\mathrm{mid}} = 15.5$~\AA\ and 28.5~\AA\ are shown in Fig.~\ref{fig:Ba45colormap}(b), further illustrating the difference between the shorter- and longer-range fits. This directly validates the previous conclusion based on our model-independent analysis that the instantaneous local structure is already symmetry-broken above the ferroelectric Curie temperature. Interestingly, a local enhancement of the tetragonal distortion also seems to be present in the multiferroic state at low temperature. Boxcar fits conducted for the other compositions showed qualitatively similar results.

As a final note on the structural characterization of \Srba, we also comment on x-ray PDF measurements that we performed with variable temperature in an \textit{in situ} magnetic field up to 5~T. No clear field dependence of any structural parameters at any temperature could be determined from the fits, indicating that if such effects are present, they are smaller than the sensitivity of our x-ray PDF measurements. This is not a surprising result, considering that antiferromagnets do not typically show a large response to an applied magnetic field.

\subsection{Characterization of the local magnetic structure}
We now present data relating to the local magnetic structure of \SBMTO, with particular emphasis on short-range magnetic correlations above \TN. We begin with magnetic PDF analysis, followed by polarized neutron scattering, and finally \muSR.

\subsubsection{Magnetic PDF analysis}
Magnetic PDF data are typically extracted from the total PDF (i.e. the sum of the nuclear and magnetic PDF signals) by subtracting the best-fit nuclear PDF from the total PDF data. The resulting fit residual contains the mPDF signal in addition to the usual components of the fit residual, i.e. noise and errors from the nuclear PDF fit. In \SBMTO, the mPDF is quite small in magnitude, comparable to the noise and errors. Therefore, we took the additional step of subtracting the nuclear PDF fit residual obtained at a high temperature ($\gtrsim 350$~K, high enough that no short-range magnetic correlations survive) from the fit residuals at all lower temperatures. This removes any temperature-independent contributions to the fit residual such as systematic errors that might otherwise swamp the mPDF signal.

\begin{figure}[h]
    \includegraphics[width=75mm]{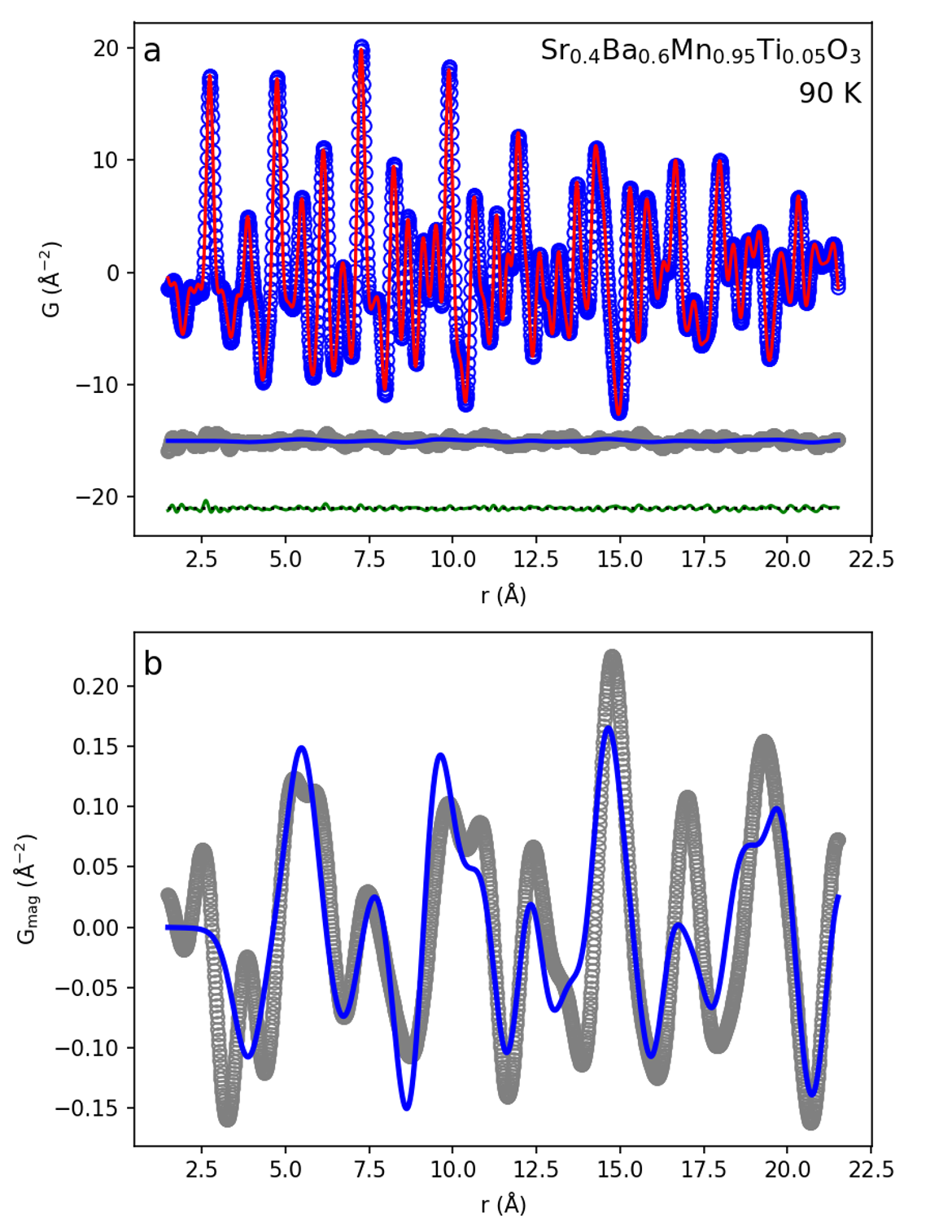}
    \caption{(a) Neutron PDF fit to \sampleB\ at 100~K. The top set of curves shows the data (blue symbols) and total PDF fit (red curve). The gray curve offset vertically below is the isolated mPDF signal, i.e. the total experimental PDF signal with the calculated nuclear PDF and the nuclear PDF fit residual from 340~K subtracted. The best-fit mPDF based on the known G-type AFM order is overlaid in blue. The overall fit residual is shown in green at the bottom of the figure. (b) Zoomed-in view of the mPDF signal (gray curve) and fit (blue curve). The experimental signal has been smoothed via a Fourier filter that removes contributions beyond 6~\AA$^{-1}$, where the magnetic scattering intensity is negligible due to the magnetic form factor.}
    \label{fig:mpdf}
\end{figure}

In Fig.~\ref{fig:mpdf}(a), we plot the total neutron PDF fit for \sampleB\ at 100~K (top set of red and blue curves), together with the isolated mPDF signal and fit (middle set of gray and blue curves).
This demonstrates the small signal weight of the mPDF relative to the total PDF. We show a close-up view of the mPDF data and fit in Fig.~\ref{fig:mpdf}(b). The fit agrees well with the data, confirming that we are sensitive to the magnetic signal in the data. Positive and negative mPDF peaks indicate predominantly parallel and antiparallel alignment between spins, respectively. The alternating negative and positive peaks in Fig.~\ref{fig:mpdf}(b) are therefore a reflection of the antiferromagnetic structure of \SBMTO.

\begin{figure}[b]
    \centering
    \includegraphics [width=75mm]{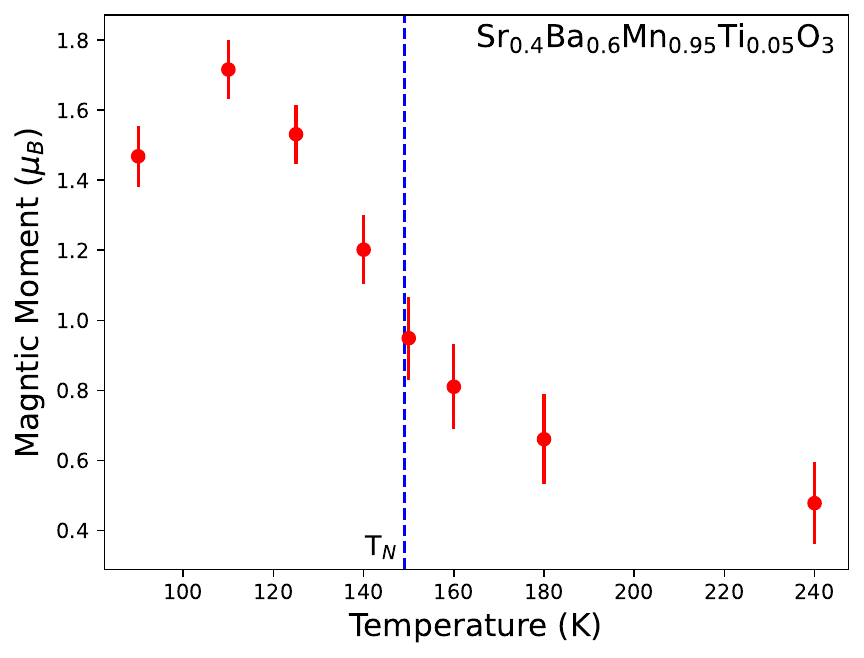}
    \caption{Locally ordered magnetic moment in \sampleB\ obtained from mPDF analysis. The persistent ordered moment above \TN\ arises from short-range AFM correlations. The anomalous drop in the ordered moment at 90~K is expected to be an artifact of imperfect thermal equilibrium.}
    \label{fig:LMOP}
\end{figure}

We performed mPDF fits to the neutron PDF data for \sampleA\ and \sampleB\ for all temperatures below 350~K, from which we extracted the local magnetic order parameter (LMOP) as a function of temperature. This should be interpreted as the magnitude of the instantaneous locally ordered moment among near-neighbor Mn$^{4+}$ spins, which will remain nonzero even above \TN\ when short-range magnetic correlations are present. We plot the LMOP for \sampleB\ in Fig.~\ref{fig:LMOP}.
The ordered moment decreases gradually with temperature, as expected, but clearly remains nonzero up to at least 240~K, which is nearly 100~K above the long-range AFM transition temperature $T_{\mathrm{N}}=149$~K (blue vertical line). Similar results are seen for \sampleA. This provides direct and quantitative evidence for short-range mannnngnetic correlations above \TN\ in \Srba.

\subsubsection{Polarized neutron scattering}
\begin{figure}
    \centering
    \includegraphics [width=75mm]{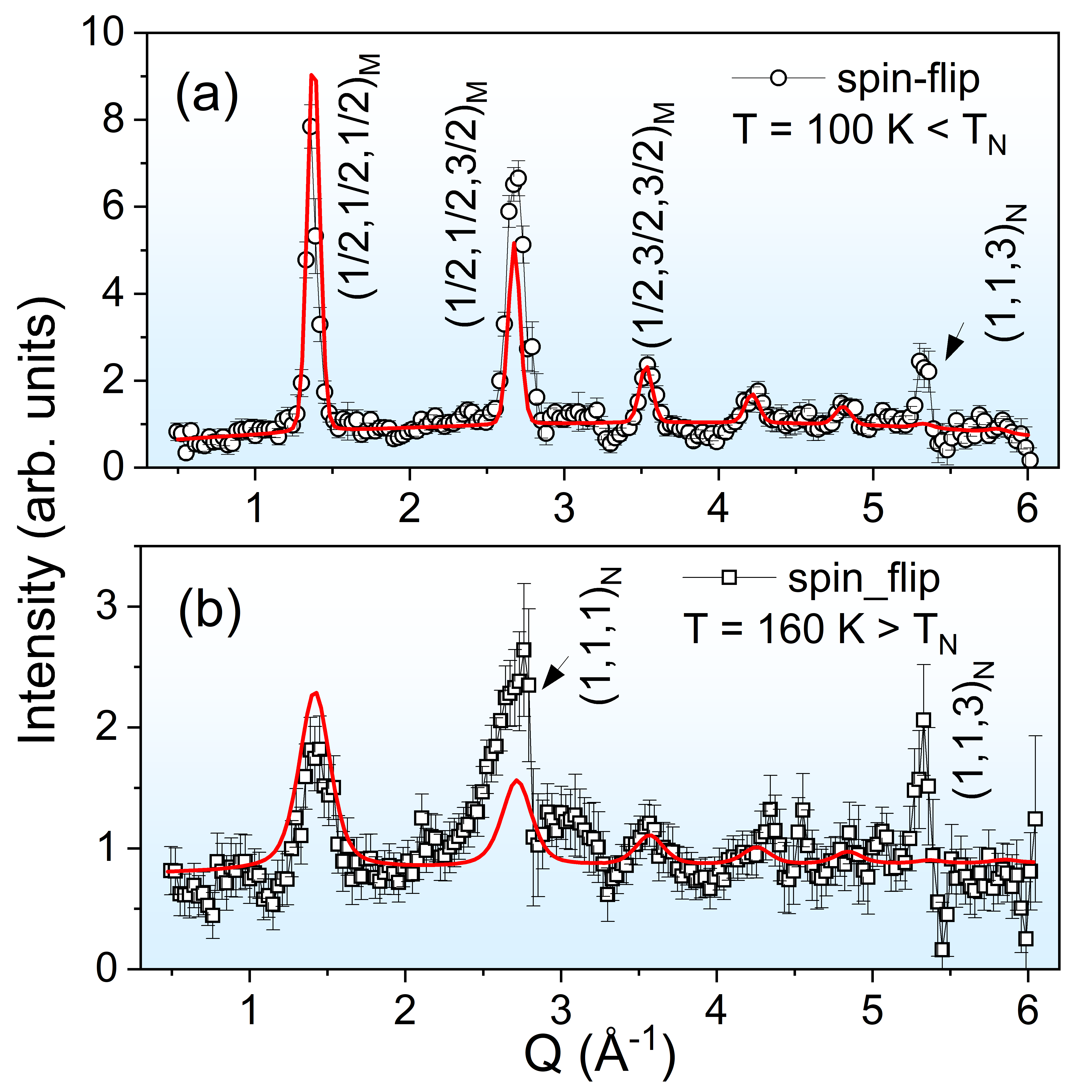}
    \caption{Spin-flip magnetic scattering from \sampleB\ at 100~K [(a); AFM state] and 160~K [(b); PM state]. The black symbols show the experimental data, while the red curves are calculated based on the model explained in the main text. Nuclear Bragg peaks causing artifacts in the magnetic signal are indicated with a subscript N.}
    \label{fig:HYSPEC}
\end{figure}

Considering the small magnitude of the mPDF signal relative to the total PDF, it is useful to confirm the presence of short-range magnetic correlations above \TN\ using polarized neutron scattering, which provides excellent sensitivity to diffuse magnetic scattering. In Fig.~\ref{fig:HYSPEC}, we plot the spin-flip (magnetic) scattering cross section of \sampleB\ at 100~K (below \TN) in panel (a) and 160~K (above \TN) in panel (b). The experimental data, shown by the black symbols, exhibit sharp magnetic Bragg peaks below \TN\ and diffuse peaks above \TN, indicative of long-range and short-range magnetic correlations, respectively. The red curves are the calculated scattering patterns using a model of the G-type AFM order consisting of long-range and short-range correlations, respectively, showing reasonable agreement with the data. The magnetic peaks at 100~K are indexed by the propagation vector $k=(1/2,1/2,1/2)$ and were modeled using the magnetic space group $I_c4cm$  in  BNS notation and $I4cm.1'_c[rP4mm]$ in the new unified (UNI) notation~\cite{campb;aca22} with corresponding lattice transformation $(a-b, a+b, 2c)$. The Mn moment is parallel to the $c$-axis and was refined to be 1.49(1)~$\mu_{\mathrm{B}}$ at 100~K. 
Quantitative agreement between the data and model  is unrealistic considering the statistical noise in the data and the leakage from large (111) nuclear Bragg peak located around 2.8~\AA$^{-1}$, which is about 60 times stronger than the magnetic scattering. Nevertheless, the presence of diffuse features in the spin-flip channel is sufficient to confirm that short-range AFM correlations survive above \TN, as expected from the mPDF analysis.

\subsubsection{Muon spin relaxation}
\begin{figure}
    \centering
    \includegraphics [width=80mm]{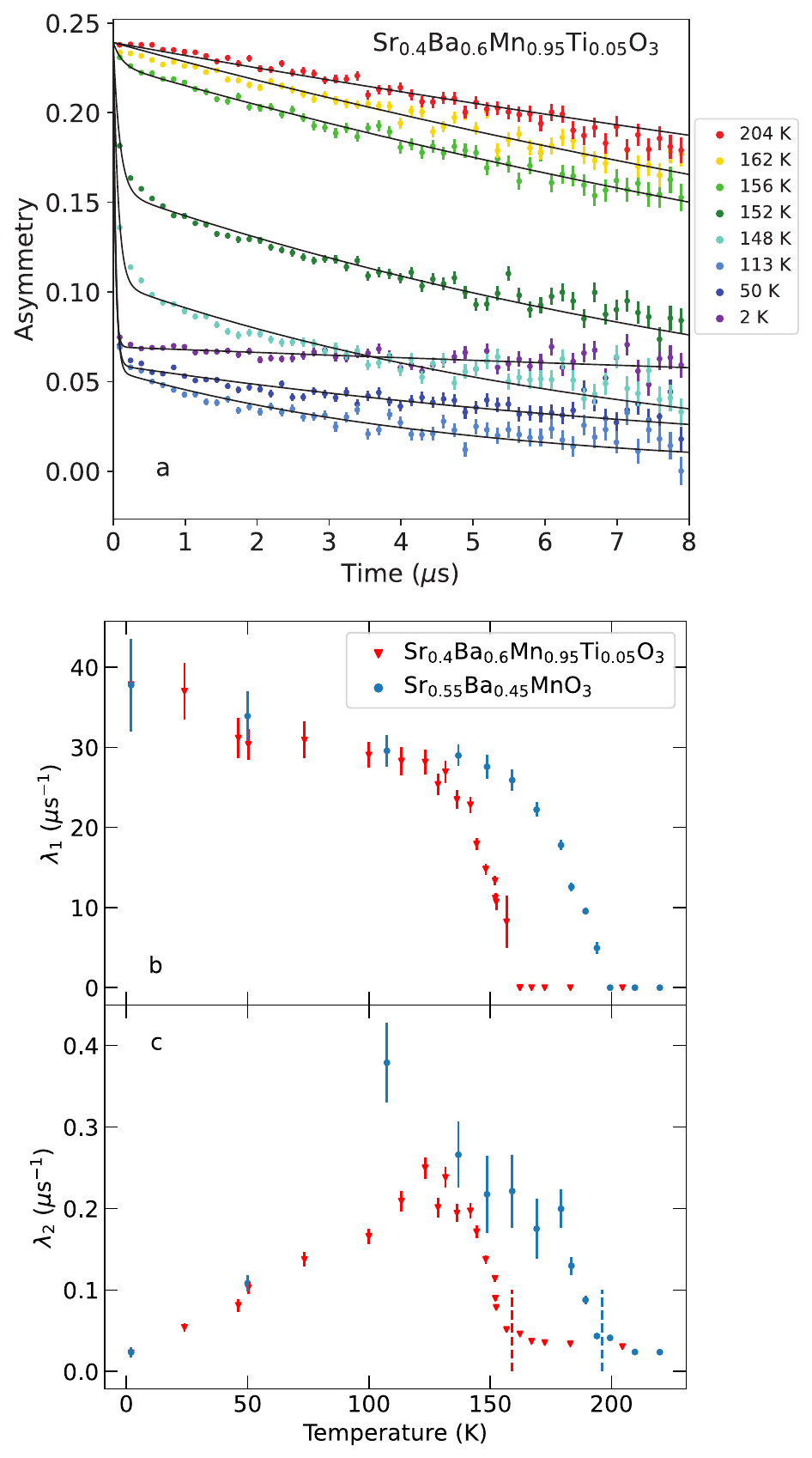}
    \caption{(a) \muSR\ time spectra for \sampleB\ at various temperatures above and below \TN. The colored symbols show the data, and the black curves are fits described in the main text. (b) Temperature-dependent relaxation rate corresponding to the fast front end of the asymmetry spectra for the samples with $(x, y) = (0.45, 0)$ and (0.6, 0.05), extracted from least-squares fits. (c) Same as for (b), but showing the relaxation rate corresponding to the long-time tail of the asymmetry spectra. The dashed vertical lines indicate \TN\ for the corresponding sample. }
    \label{fig:musr}
\end{figure}

As an independent verification of the local magnetic properties of \Srba\ revealed by neutron scattering, we performed \muSR\ measurements on \sampleA\ and \sampleB. Asymmetry spectra collected at various temperatures across the antiferromagnetic transition are shown for \sampleB\ in Fig.~\ref{fig:musr}(a).
Above \TN, the asymmetry relaxes slowly, with a single exponentially decaying component. Below the transition, we observe a fast front-end relaxation and a slower long-time component with amplitude about one-third the initial asymmetry amplitude of $\sim$0.24, as expected for magnetically ordered powder samples. The transition region is evident between 148 and 156~K for \sampleB\ in Fig.~\ref{fig:musr}(a), consistent with the transition temperature of 149~K determined from magnetometry. %This discrepancy may be due to imperfect temperature calibration and/or the influence of short-range order above \TN\ on the \muSR\ spectra.
Close inspection of the spectra shows that the relaxation in the paramagnetic state becomes more rapid as the temperature approaches \TN\ from above. This is a signature of critical spin dynamics near the transition, consistent with the development of short-range magnetic correlations observed by mPDF and polarized neutron scattering. The asymmetry spectra for \sampleA\ likewise confirm a bulk magnetic transition at 195~K and show the same dynamic behavior just above \TN.

We analyzed the \muSR\ data more quantitatively by performing fits to the asymmetry spectra using the program BEAMS~\cite{peter;gh21}. We modeled the data using a sum of two exponential components, one for the fast front end and the other for the slower, long-time relaxation. The total initial amplitude of the two components was constrained to be equal to a constant representing the total initial asymmetry determined by a global fit to all the spectra, found to be 0.2390(3) for \sampleA\ and 0.2356(4) for \sampleB. Above \TN, the fast front-end component was fixed to zero amplitude. The fast and slow relaxation rates, labeled $\lambda_1$ and $\lambda_2$, respectively, are shown in Fig.~\ref{fig:musr}(b) and (c). $\lambda_1$ is proportional to the static field at the muon site~\cite{uemur;ms99} and therefore serves as an order parameter for the magnetic transition, revealing a rather continuous evolution of the static field for the two samples, as seen in Fig.~\ref{fig:musr}(b). The slower rate $\lambda_2$ is related to the magnetic fluctuations~\cite{uemur;ms99}. The slight but observable increase in $\lambda_2$ as the temperature approaches \TN\ from above confirms the development of short-range magnetic correlations mentioned previously. The broad peak in $\lambda_2$ below \TN\ is evidence of critical spin dynamics, with the decrease at lower temperature caused by the freezing out of magnetic excitations.

\section{Discussion and Conclusion}
Based on the total scattering, polarized neutron scattering, and \muSR\ data presented here, the local electronic environment in \SBMTO\ is characterized by widespread local symmetry breaking. This includes not only short-range tetragonal distortions resulting in local electric dipole formation deep into the high-temperature paraelectric phase, but also short-range AFM correlations that survive well above \TN. The occurrence of local symmetry breaking above ferroelectric or magnetic transitions is by no means unique or even particularly uncommon, but these observations nevertheless confirm the relevance of such behavior for \SBMTO\ and possibly other multiferroic systems, where local symmetry breaking has received mostly limited attention in the past.

\begin{figure}
    \centering
    \includegraphics[width=75mm]{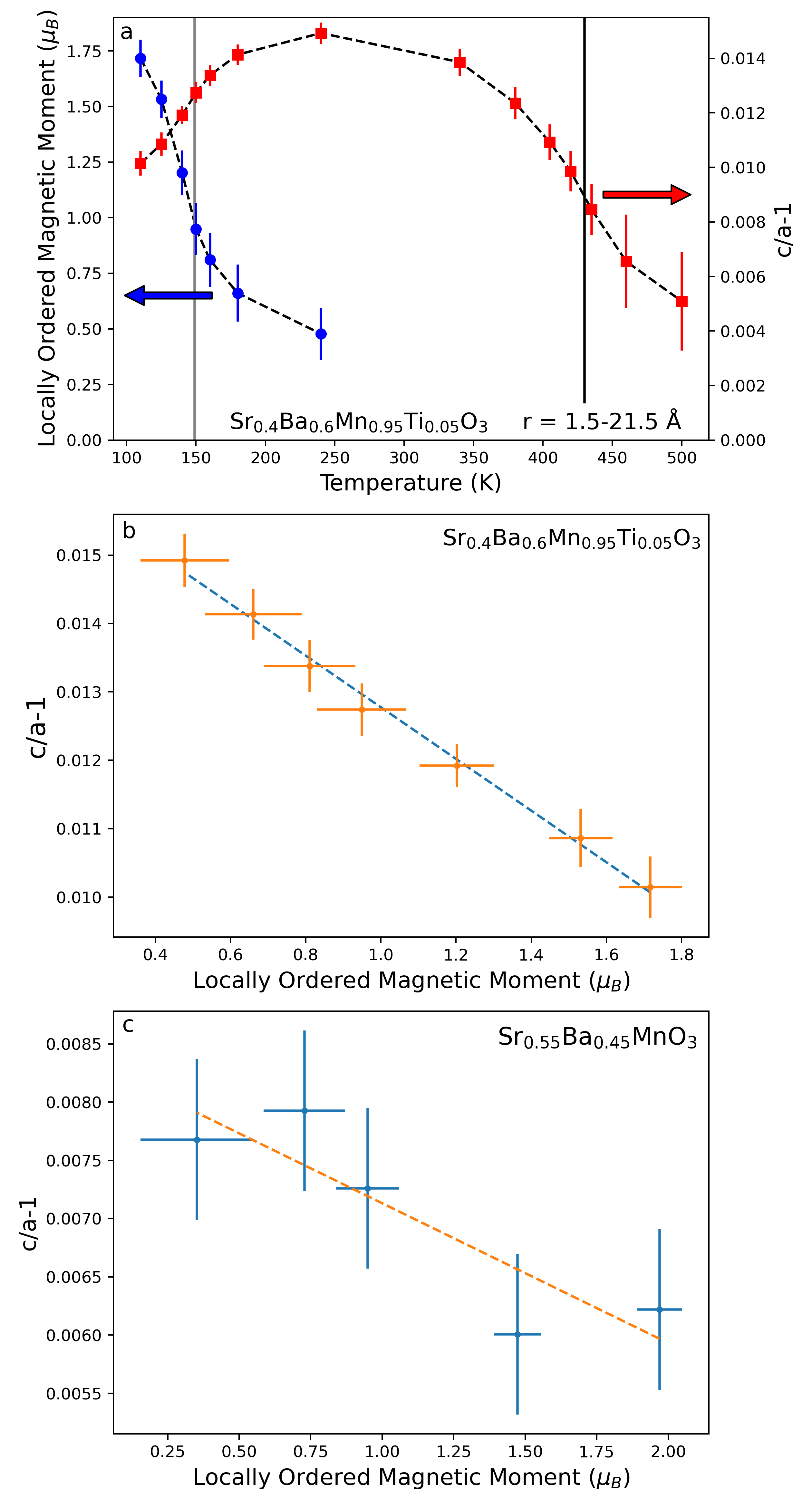}
    \caption{(a) Temperature dependence of the locally ordered magnetic moment for \sampleB\ (left vertical axis) and the tetragonal distortion $c/a-1$ for fitting range of 1.5 -- 21.5~\AA\ (right vertical axis). (b) Tetragonal distortion versus local magnetic order parameter for \sampleB, highlighting spontaneous linear magnetoelectric coupling in the multiferroic state. (c) Same as (b), but for \sampleA.}
    \label{fig:overlay}
\end{figure}

Considering the behavior of the local atomic and magnetic structure together allows us to interrogate the coupling between the electric and magnetic degrees of freedom, a central issue for multiferroics. To facilitate this, we plot in Fig.~\ref{fig:overlay}(a) the magnitude of the tetragonal distortion $c/a-1$ in \sampleB\ as a function of temperature for a short PDF fitting range, together with the local magnetic order parameter obtained from the mPDF fits.

An issue not fully resolved in previous studies is the reason for the reduction of the tetragonal distortion upon approaching the AFM transition from above, beginning around 250~K in \sampleB. Here, it is clear that this reduction coincides precisely with the onset of short-range AFM correlations above \TN, providing direct evidence that the magnetic fluctuations are responsible for the reduced tetragonal distortion.

To illustrate this further, we plot the tetragonal distortion against the local magnetic order parameter in Fig.~\ref{fig:overlay}(b). The relationship can be well described by a line with slope -0.0037(1)~$\mu_{\mathrm{B}}^{-1}$, demonstrating spontaneous linear magnetoelectric coupling in \SBMTO. We emphasize that this linear relationship extends even to the short-range magnetic correlations above \TN, confirming that short-range correlations can facilitate the magnetoelectric coupling characteristic of the multiferroic phase. This coupling could be achieved either directly through interactions between the electric dipoles and the fluctuating spins or indirectly via magnetostructural coupling, by which the magnetic correlations drive a response in the structure, which then in turn influences the magnitude of the ferroelectric distortion. Further investigation, including theoretical work to model the microscopic interactions in \SBMTO, will likely be necessary to answer this question. An equivalent plot for \sampleA\ is shown in Fig.~\ref{fig:overlay}(c). The same trend appears, although it is noisier due to the smaller structural distortion in \sampleA\ and fewer temperature points for which data were collected.

We note that the phenomenon of short-range magnetism driving long-range changes in the average structure, such as that seen in \SBMTO, has also been observed in a diversity of other systems, such as a geometrically frustrated magnet~\cite{frand;prb20} and a giant magnetostrictive material~\cite{baral;afm23}. More generally, these results highlight a recurring theme in complex materials with intertwined orders: focusing only on the average, long-range behavior of the relevant degrees of freedom (e.g. lattice, charge, and spin) may result in an incomplete view of the system. Deeper understanding often requires careful study of the short-range, local correlations within and between these degrees of freedom, as revealed by local probes such as PDF and \muSR\ used here.

\textbf{Acknowledgements}

We thank Milinda Abeykoon for help with the x-ray PDF experiment, along with Michelle Everett, Jue Liu, and Cheng Li for help with the neutron PDF experiment. For assistance with the \muSR\ experiment, we thank Gerald Morris, Shan Wu, Shannon Haley, and Emma Zappala. B.A.F. and B.J. were supported by the U.S. Department of Energy, Office of Science, Basic Energy Sciences (DOE-BES) through Award No. DE-SC0021134. C.Z.S. was supported by the College of Physical and Mathematical Sciences at Brigham Young University. Work in the Materials Science Division at Argonne National Laboratory (materials synthesis, magnetic characterization and preliminary x-ray diffraction) was supported by the U.S. Department of Energy, Office of Science, Basic Energy Sciences, Materials Science and Engineering Division. This research used beamline 28-ID-1 of the National Synchrotron Light Source II, a U.S. Department of Energy (DOE) Office of Science User Facility operated for the DOE Office of Science by Brookhaven National Laboratory under Contract No. DE-SC0012704. This study used resources at the Spallation Neutron Source (SNS), a DOE Office of Science User Facility operated by the Oak Ridge National Laboratory.

\end{document}